\documentclass[
reprint,
showpacs,
preprintnumbers,
nofootinbib,
amsmath,
amssymb,
aps,
prd,
]{revtex4-1}
\usepackage{enumerate}
\usepackage{color}
\usepackage{graphicx}
\usepackage{dcolumn}
\usepackage{bm}
\usepackage{empheq}
\usepackage{nomencl}
\usepackage{hyperref}
\usepackage{mathrsfs}
\usepackage{amsmath}
\usepackage{url}
\usepackage{epstopdf}
\hypersetup
{
	colorlinks,%
	citecolor=blue,%
	linkcolor=magenta,%
	urlcolor=blue,%
}

\usepackage{subfigure}

\begin{document}
	\newcommand{\red}[1]{\textcolor{red}{#1}}
	\newcommand{\green}[1]{\textcolor{green}{#1}}
	\newcommand{\blue}[1]{\textcolor{blue}{#1}}
	\newcommand{\cyan}[1]{\textcolor{cyan}{#1}}
	\newcommand{\purple}[1]{\textcolor{purple}{#1}}
	\newcommand{\yellowbox}[1]{\colorbox{yellow}{#1}}
	\newcommand{\purplebox}[1]{\colorbox{purple}{#1}}
	\newcommand{\yellow}[1]{\textcolor{yellow!70!red}{#1}}
	\newcommand{\sch}{Schwarzschild}
	\newcommand{\be}{\begin{equation}}
		\newcommand{\en}{\end{equation}}
	\title{True gravitational atom: Spherical cloud of dilatonic black holes}
	\author{Yang Huang $^1$}\email{sps\_huangy@ujn.edu.cn}
	\author{Hongsheng Zhang $^{1}$}\email{sps\_zhanghs@ujn.edu.cn}
	\affiliation{
		$^1$ School of Physics and Technology, University of Jinan, 336, West Road of Nan Xinzhuang, Jinan 250022, Shandong, China}

	\begin{abstract}
		Black hole as elementary particle is a fairly glamorous idea. For ordinary black holes, a surrounding particle inevitably penetrates into the interior of the hole since the center of the hole is an infinite potential well. For the first time we demonstrate that an extreme dilatonic black hole in spherical symmetry perfectly behaves as an atom, in the sense that its surrounding cloud of particles are completely stable. Thus we reach a spherical cloud of dilatonic black hole. We find exact wave functions of the cloud for arbitrary gravitational fine structure constant $\mu M$, and clear the underlying physical nature of the stability.  Through careful studies of  the exact wave function, we find the spectrum of this system. We discuss the physical meaning of this discovery especially from considerations of entropy,  and the resultant possibility to explore quantization of gravitational waves from coming observations.

	\end{abstract}
	
	\maketitle
	
	\section{Introduction}
	Black hole is the most pure object in gravity, which may be treated as atom in gravity theory. Because of the uniqueness theorem, a black hole looks like an atom even more, in the original sense of this word. Direct quantization of black hole itself runs up against stone walls again and again. Actually even to propose a feasible test for quantization of stellar-mass or super/hyper massive black holes is very difficult.  In view of the construction history of quantum mechanics, one may first find the quantized energy level of a surrounding particle controlled by gravity force, and then explore the properties of the
	gravitational waves triggered by transitions between energy levels. If one finds absorption lines in the stochastic gravitational waves background and identifies the source of this line, one may make a real step towards quantizations of gravitational waves.
	
	To find a black hole with stable energy level structure similar to an atom is not a trivial work. Generally, a black hole has interior, which is different from an elementary particle. For a surrounding particle, the center of the interior of a hole is an infinite potential well. Thus, any matter wave (particle) surrounding the hole will leak into the hole through quantum tunnelling, which causes that the surrounding waves are unstable \cite{Detweiler:1980uk,PhysRevD.99.123011}. Mathematically, the eigenvalue of the frequency is a complex number, in which the imaginary part implies that the surrounding particle is decaying. It seems
	rather difficult to evade such a situation. For a rotating hole, superradiance process tells that a particle with particular frequency can excite more particles with exactly same frequency \cite{Starobinsky:1973aij,PhysRevD.10.3194,Brito:2015oca}.
	If the two processes, leakage and superradiance processes, balance each other, the surrounding particles may construct a cloud \cite{PhysRevD.86.104026,Huang:2016qnk,Huang:2017whw}.
	If the backreaction of the cloud to the metric is considered, one arrives at a hairy black hole \cite{Herdeiro:2014goa}. Such a cloudy/hairy hole only could be unstable against specific modes of perturbations of the metric \cite{PhysRevLett.120.171101}. Real stable energy level structure like a Hydrogen atom, which does not depends on such a delicate balance, is critically wanted for studies of quantization of gravitational waves.
	
	From other aspects, a black hole does not seem to behave like an elementary particle. Generally a black hole has a large entropy, which indicates that it has a large amount of internal degrees of freedom. This is rather different from an elementary particle. Based on thermodynamical considerations and fluctuation analysis, an interesting conjecture is proposed in \cite{Holzhey:1991bx,doi:10.1142/S0217732391002773}
	that extreme dilatonic black hole behaves as elementary particle. Put simply, the area of event horizon of a dilatonic black hole goes to zero at the extreme limit. Thus one deduces that its entropy vanishes for an extreme black hole from thermodynamic reasoning. One has to arrive at the conclusion that an extreme dilatonic black hole has no internal degree of freedom if statistical mechanic principle works well for such a hole. Through rigorous studies of scalar wave surrounding an extreme dilatonic black hole, we find the particles in bound states are stable, which do not penetrate into the interior of the hole. This presents strong evidence that this hole has no internal degree of freedom.
	
	This paper is organized as follows. In Sec.\ref{Sec: Quasibound states}, we briefly review the quasibound states of a massive scalar field for dilatonic black holes, and show that the quasibound states tend to be stable in the extremal limit. Next, in Sec.\ref{Sec: Bound states}, we construct the stable solutions of a massive scalar field for exactly extreme hole. Finally, we conclude this paper in Sec.\ref{Sec: conclusion}. We use the units $G=\hbar=c=1$.
	
	\section{Quasibound states of a massive scalar field}\label{Sec: Quasibound states}
	\subsection{Radial equation and effective potential}
	
	Dilaton gravity, or called scalar-tensor theory, as one the most significant modified gravity,  takes a special status in gravity theories. It can be traced back to Kaluza-Klein compactification \cite{Appelquist:1987nr} and Dirac's large number hypothesis \cite{Dirac:1937ti}. In either case, an extra non-minimally coupled scalar field is introduced in the effective field equation. Dilaton gravity is the minimal modification, and a strong  competitor, of general relativity, in which only one extra freedom, the dilaton field, is involved in gravity interaction. Many different modified gravity theories can be reduced dilaton gravity in low energy limit, though the physical starting points of them may be completely different. A recent example is string theory. Heterotic string theory reduces to a dilaton gravity in low energy limit, the action is
	\begin{equation}
		S=\int d^4x\sqrt{-g}\left[R-2\left(\nabla\phi\right)^2-e^{-2\phi}F_{\mu\nu}F^{\mu\nu}\right],
	\end{equation}
	where $\phi$ is the dilaton field, and $F_{\mu\nu}=\partial_\mu A_\nu-\partial_\nu A_\mu$ is the electromagnetic tensor. In Refs.\cite{Garfinkle:1990qj,Gibbons:1987ps}, a spherical black hole solution was derived by Gibbons, Maeda, Garfinkle, Horowitz, and Strominger (GMGHS). In Einstein frame, the metric of the GMGHS black hole reads,
	\begin{equation}\label{Eq: metric}
		ds^2=-Fdt^2+F^{-1}dr^2+r^2G\left(d\vartheta^2+\sin^2\vartheta d\varphi^2\right),
	\end{equation}
	with
	\begin{equation}
		F(r)=1-\frac{2M}{r},\;\text{and}\;G(r)=1-\frac{Q^2}{Mr},
	\end{equation}
	where $M$ and $Q$ are the mass and charge of the black hole, respectively.
	The GMGHS solution has one event horizon at $r_+=2M$, and a singular surface at $r_-=Q^2/M$.
	For $Q<Q_{\mathrm{max}}\equiv\sqrt{2}M$, the singularity is enclosed by the event horizon.
	In the extreme case $Q=Q_{\mathrm{max}}$, and the singularity coincides with the horizon.
	The surface gravity at the event horizon reads,
	\begin{equation}
		\kappa=\frac{1}{2r_+},
	\end{equation}
	which directly follows the case of \sch. The Wald entropy becomes different,
	\begin{equation}
		S=\pi  r_+^2\left(1-\frac{r_-}{r_+}\right).
	\end{equation}
	By using $Q^2=r_+r_-/2$, one confirms that the GMGHS black hole satisfies the first law. The critical point is that the entropy vanishes when the hole becomes an extreme one, which leads to an inescapable conclusion that
	the extreme black hole has no internal degree of freedom if the principle of statistical physics works soundly. This property is exactly in analogy with an elementary particle. Thus, a reasonable conjecture is that the surrounding
	matter wave cannot permeate into the interior of the extreme GMGHS black hole, since there is no phase space to accommodate the  permeated waves. As a contrast, the entropy of extreme RN black hole does not vanish. So a
	surrounding particle has probability to fall into it. We shall demonstrate that bound state of extreme GMGHS black hole is really stable, as expected.

	For convenience, we introduce a normalized charge $q=Q/Q_{\mathrm{max}}\in\left[0,1\right]$, and parameterize the charge by \begin{equation}\label{Eq: eta}
		q=1-e^{-\eta}.
	\end{equation}
	The Schwarzschild black hole corresponds to $\eta=0\;(q=0)$, while the extreme GMGHS black hole corresponds to $\eta\rightarrow\infty$ $(q\rightarrow1)$.
	
\begin{figure}
		\centering	
		\includegraphics[width=0.45\textwidth,height=0.35\textwidth]{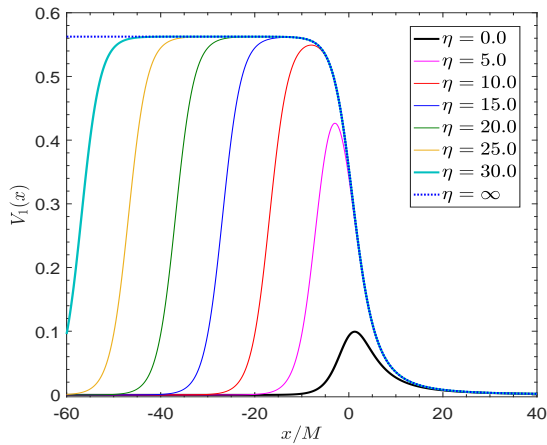}		
		\caption{Comparison of the effective potential for different $\eta$, with $l=1$ and $\mu M=0$. Here, $x$ is the tortoise coordinate defined below Eq.(\ref{Eq: the radial eq}).}
		\label{Fig: potential}
	\end{figure}

	Scattering states of a massless scalar field by the GMGHS black hole has been studied in \cite{Huang:2020bdf}. Here, we investigate the bound states of a massive scalar field $\Psi$, which is different from the dilaton field in the theory. In the present study we work in probe limit, that is, $\Psi$ has no back reaction to the spacetime metric. The Klein-Gordon equation of a scalar field
	with mass $\mu$ in GMGHS spacetime is given by $\nabla_\mu\nabla^\mu\Psi=\mu^2\Psi$.
	This equation admits the separable solutions of the form \cite{PhysRevD.103.044062}
	\begin{equation}\label{Eq: separate variable}
		\Psi=e^{-i\omega t}\frac{\psi(r)}{r\sqrt{G(r)}}Y_{lm}(\vartheta,\varphi),
	\end{equation}
	where the radial function $\psi$ obeys the radial equation
	\begin{equation}\label{Eq: the radial eq}
		\frac{d^2\psi}{dx^2}+\left[\omega^2-V_l(r)\right]\psi=0,
	\end{equation}
	with $x=\int dr/F$ the tortoise coordinate, and the effective potential given by
	\begin{equation}\label{Eq: effective potential}
		\begin{aligned}
			V_l(r)=&\frac{F(r)}{G(r)}\left[\frac{F'(r)}{r}+\frac{l(l+1)}{r^2}+\mu^2G(r)\right]\\&-\frac{2M^2q^2}{r^4}\frac{F(r)}{G(r)^2}\left[1+\frac{q^2}{2}\left(1-\frac{6M}{r}\right)\right].
		\end{aligned}
	\end{equation}
	Fig. \ref{Fig: potential} compares $V_l$ for different values of $\eta$.
	As pointed out in \cite{PhysRevD.103.044062}, the width of the potential barrier increases monotonously with increasing $\eta$.
	In the extreme limit $\eta\rightarrow\infty$, the width of the potential barrier increases without bound, and the black hole absorption is heavily suppressed. This property is crucial for the existence of long-lived modes
	of massive scalar field adhered to a GMGHS black hole.
	
	\subsection{Quasibound state spectrum}
	The penetration of the particle into the interior of the black hole is confirmed by studies of energy flux of the particles at quasibound state, as follows. One clearly observes that energy flows into the black hole by explorations of energy flux in the black hole spacetime. Quasibound states are solutions of Eq.(\ref{Eq: the radial eq}) that satisfy the ingoing wave condition at the horizon and decrease exponentially at infinity. The quasibound state spectrum was studied in \cite{PhysRevD.103.044062}.
	It was shown that in the near extreme limit $\eta\gg1$, the quasibound state frequencies can be obtained by matching the near horizon and far field solutions. Due to the black hole absorption, the eigenfrequency is generally a complex number (expressed by $\omega=\omega_R+i\omega_I$) with a negative imaginary part. Both the real and imaginary parts of the frequency are tuned by the gravitational fine structure constant $\alpha\equiv\mu M$. For $\alpha<L/2$, the real part of the frequency is given by
	\begin{equation}
		\frac{\omega_R}{\mu}\approx 1-\frac{\alpha^2}{2\tilde{n}^2}-\frac{2(1+\tau)\alpha^4}{\tilde{n}^3L}+\frac{(\tau+15/8)\alpha^4}{\tilde{n}^4}+\cdots,
	\end{equation}
	where $L=l+1/2$, $\tilde{n}=n+l+1$ with $n=0,1,2,\cdots$, and $\tau=1-q^2$ is a small number for near extreme GMGHS black holes. The imaginary part is given by
	\begin{equation}
		\omega_I\propto-e^{-2\tilde{\beta}\eta},\;\;\;\text{with}\;\;\;\tilde{\beta}=\sqrt{L^2-4\omega_R^2}.
	\end{equation}
	As mentioned above, for $\eta\gg1$, $\omega_I$ tends to zero exponentially since the black hole absorption is suppressed by a large potential barrier.
	
	
	\subsection{Energy and flux}
	The Kerr-Schild coordinates $\tilde{x}^{\mu}=\left[\tilde{t},r,\vartheta,\varphi\right]$ is adopted to study the black hole absorption of the scalar field, where the new time coordinate is given by \cite{Degollado:2013bha}
	\begin{equation}
		\tilde{t}=t+x-r,
	\end{equation}
	where $x$ is the tortoise coordinate defined below Eq.(\ref{Eq: the radial eq}). Then, the GMGHS metric (\ref{Eq: metric}) becomes
	\begin{equation}
		\begin{aligned}
			ds^2=&\tilde{g}_{\mu\nu}d\tilde{x}^{\mu}d\tilde{x}^{\nu}\\
			=&-Fd\tilde{t}^2+(2-F)dr^2+2(1-F)d\tilde{t}dr+r^2Gd\Omega^2.
		\end{aligned}
	\end{equation}
	For complex massive scalar fields, the energy-momentum tensor is given by
	\begin{equation}\label{Eq: energy tensor}
		T_{\mu\nu}=2\Psi^*_{,(\mu}\Psi_{,\nu)}-g_{\mu\nu}\left(\Psi^*_{,\gamma}\Psi^{,\gamma}+\mu^2|\Psi|^2\right),
	\end{equation}
	where $X_{(\mu}Y_{\nu)}\equiv\left(X_{\mu}Y_{\nu}+X_{\nu}Y_{\mu}\right)/2$. The timelike Killing vector field $k=\partial_t$ defines a conserved current $J^{\mu}=-T^{\mu}_{\;\;\nu}\xi^{\nu}$, where
	$\xi^{\mu}=\left[1,0,0,0\right]$. Using the conservation law $\nabla_\mu J^\mu=0$ and Gauss's theorem, we obtain
	\begin{equation}
		\frac{dE}{d\tilde{t}}=-\mathcal{F},
	\end{equation}
	where
	\begin{equation}\label{Eq: Def of E}
		E=\int_{\Sigma_{\tilde{t}}} d^3x\sqrt{-g}J^0,
	\end{equation}
	is the total energy of the scalar field, and $\Sigma_{\tilde{t}}$ is a spacelike hypersurface with $\tilde{t}=0$. Flux at $r=r_+$ is given by
	\begin{equation}
		\mathcal{F}=\int_{\partial\Sigma_{\tilde{t}}}r^2G T^{1}_{\;\;0}\sin\vartheta d\vartheta d\varphi,
	\end{equation}
	where $\partial\Sigma_{\tilde{t}}$ is the boundary of the hypersurface, i.e., $\tilde{t}=0$, $r=r_+$, $\vartheta\in\left[0,\pi\right]$, and $\varphi\in\left[0,2\pi\right]$. If we decompose the scalar field as $\Psi(\tilde{t},r,\vartheta,\varphi)=\sum_{lm}\phi_l(\tilde{t},r)Y_{lm}(\vartheta,\varphi)$, then, the flux can also be decomposed as a sum over all angular modes
	\begin{equation}
		\mathcal{F}=2r^2G\sum_{l}\mathcal{F}_l
	\end{equation}
	where
	\begin{equation}
		\mathcal{F}_l=\tilde{g}^{01}\big|\partial_{\tilde{t}}\phi_l\big|^2+\tilde{g}^{11}\text{Re}\left(\partial_{\tilde{t}}\phi_l^*\partial_r\phi_l\right),
	\end{equation}
	Clearly, the flux is always positive for non-extreme GMGHS black hole.
	Therefore, the energy $E$ decays with time due to the black hole absorption. For extreme GMGHS black hole, however, the flux vanishes at the horizon since $G(r_+)=0$. This implies that in this case, the energy of the
	scalar field is conserved. No energy penetrates into the interior of the hole. Thus, stationary states may exist in extreme GMGHS spacetime.
	
	\section{Bound states in extremal GMGHS spacetime}\label{Sec: Bound states}
	\subsection{Bound state spectrum}
	Based on the discussion above, we expect that in the extreme limit, real bound states with $\omega_I=0$ exist. They are stable, and thus the black hole becomes a true gravitational atom without decaying.
	
\begin{figure}
		\centering	
		\includegraphics[width=0.45\textwidth,height=0.35\textwidth]{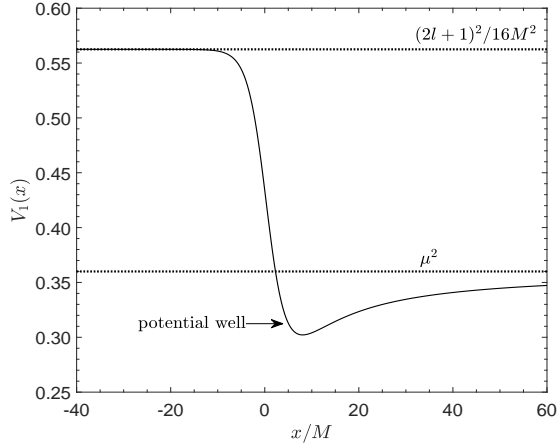}		
		\caption{Effective potential for $l=1$ and $\mu M=0.6$.}
		\label{Fig: potential extreme}
	\end{figure}

\begin{figure}
		\centering	
		\includegraphics[width=0.45\textwidth,height=0.35\textwidth]{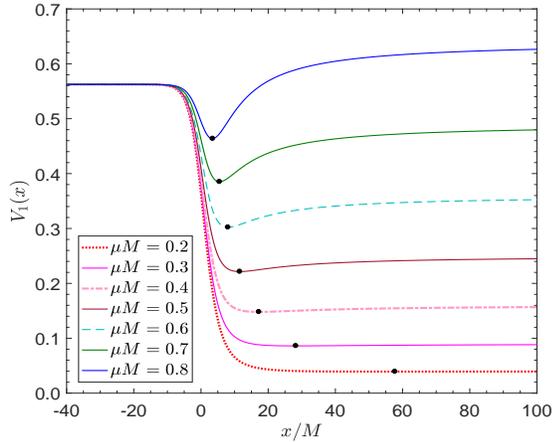}		
		\caption{Comparison of the effective potentials of $l=1$ state for different values of $\mu M$. The black points mark the local minimum of potential well.}
		\label{Fig: potential extreme massive}
	\end{figure}

	For extreme GMGHS black hole, the effective potential given in Eq.(\ref{Eq: effective potential}) becomes
	\begin{equation}\label{Eq: effective potential extreme}
		V_l(r)=\frac{l(l+1)}{r^2}+\frac{M}{r^3}\left(2-\frac{3M}{r}\right)+F(r)\mu^2.
	\end{equation}
	Unlike the non-extreme case, $V_{l}$ does not vanish at the horizon, but tends to $(2l+1)^2/16M^2$, see Fig.\ref{Fig: potential extreme}. We also observe that for $\mu M\neq0$, $V_l$ develops a potential well between the potential barrier and infinity.
	Fig. \ref{Fig: potential extreme massive} compares the effective potentials for different $\mu M$. We see that the local minimum of the potential well moves far away from the black hole with the decrease of $\mu M$.
	
	The walls on both sides of the potential well are infinitely wide.
	Thus, we choose the boundary conditions $\psi_{\omega l}(r_+)=\psi_{\omega l}(\infty)=0$.
	It is convenient to introduce the following dimensionless quantities
	\begin{equation}\label{Eq: dimensionless coord}
		z=\frac{r-r_+}{r_+};\;\;\;\epsilon=\omega M;\;\;\;\alpha=\mu M.
	\end{equation}
	Substituting Eq.(\ref{Eq: effective potential extreme}) into Eq.(\ref{Eq: the radial eq}), the radial equation becomes
	\begin{equation}\label{Eq: the radial eq z}
		\frac{d^2\psi}{dz^2}+\frac{1}{z(z+1)}\frac{d\psi}{dz}-U\psi=0,
	\end{equation}
	where
	\begin{equation}
		\begin{aligned}
			U(z)=&\frac{4(1+z)}{z}\left[\alpha^2-\left(1+\frac{1}{z}\right)\epsilon^2\right]+\frac{l(l+1)}{z^2}\\
			&+\frac{1+4z}{4z^2(1+z)^2}.
		\end{aligned}
	\end{equation}
	We find that solutions of Eq.(\ref{Eq: the radial eq z}) obeying boundary condition $\psi(z=0)=0$ can be expressed in terms of the generalized Laguerre function
	\begin{equation}\label{Eq: radial func}
		\psi(z)=z^\beta(1+z)^{1/2}e^{-kz}L_{-1/2-\beta+\kappa}^{2\beta}(2kz),
	\end{equation}
	where $k=2\sqrt{\alpha^2-\epsilon^2},\;\kappa=(4\epsilon^2-k^2)/(2k)$ and
	\begin{equation}\label{Eq: Def of beta}
		\beta=\sqrt{L^2-4\epsilon^2},\;\;\;\text{with}\;\;\;L\equiv l+1/2.
	\end{equation}
	Then, using the property of the generalized Laguerre function and boundary condition $\psi(\infty)=0$, we have
	\begin{equation}\label{Eq: matching condition}
		-\frac{1}{2}-\beta+\kappa=n,
	\end{equation}
	where $n=0,1,2,\cdots$ is the overtone. Amazingly, this is exactly the extreme limit of the matching condition of the quasibound states in the non-extreme case, see \cite{PhysRevD.103.044062} for more details. The physical nature of this bound state roots in the property of potential. The potential becomes infinitely wide in the extreme limit, which obstructs any possible penetration of the matter wave into the hole.
	
	Equation (\ref{Eq: matching condition}) determines the bound state frequencies labeled by the overtone $n$.
	From Fig. \ref{Fig: potential extreme}, the bound state frequency must obey
	\begin{equation}
		\omega_n<\text{min}\left\lbrace\mu,(2l+1)/4M\right\rbrace,
	\end{equation}
	or equivalently, $\epsilon_n=M\omega_{n}<\text{min}\left\lbrace \alpha,L/2\right\rbrace $. Following \cite{PhysRevD.103.044062}, we solve Eq.(\ref{Eq: matching condition}) with the series expansion
	\begin{equation}\label{Eq: bound state spectrum}
		\epsilon_n=\alpha\left[1+\sum_{i=1}^{\infty}C_i\alpha^{2i}\right].
	\end{equation}
	The coefficients $C_i$ can be obtained by substituting the series expansion into Eq.(\ref{Eq: matching condition}), and solving the equation order by order. Here we list the first third $C_i$:
	\begin{subequations}\label{Eq: Ci}
		\begin{empheq}{align}
			C_1=&-\frac{1}{2\tilde{n}^2},\\
			C_2=&-\frac{2}{\tilde{n}^3L}+\frac{15}{8\tilde{n}^4},\\
			C_3=&-\frac{2}{\tilde{n}^3L^3}-\frac{6}{\tilde{n}^4L^2}+\frac{17}{\tilde{n}^5L}-\frac{145}{16\tilde{n}^6},
		\end{empheq}
	\end{subequations}
	where $\tilde{n}=n+l+1$ is the principal quantum number. In practice, we have computed the coefficients up to $C_9$. But their expressions are too cumbersome to be presented here.
	
	It should be noted that $\epsilon_n$ given in Eq.(\ref{Eq: bound state spectrum}) are purely real frequencies, and they are exactly the extreme limit of the quasibound state frequencies in the non-extreme case. Therefore, the bound states obtained in this work are stable as the wave function of the scalar field will not grow or decay with time.
	
	To find possible signs of the bound states from gravitational wave detectors, we make a numerical approximation of the energy levels. In the limit of $\alpha\ll L$, the bound state has a hydrogen-like spectrum
	\begin{equation}
		\omega_{\tilde{n}}\simeq\mu\left(1-\frac{\alpha^2}{2\tilde{n}^2}\right).
	\end{equation}
	Like the photoelectric effects, gravitational wave with proper frequency will trigger transitions between energy levels. The frequency of the required graviton to trigger this process ($\tilde{m}<\tilde{n}$) is given by
	\begin{equation}
		f_{\tilde{n}\tilde{m}}\simeq\frac{\mu}{2}\alpha^2\left(\frac{1}{\tilde{m}^2}-\frac{1}{\tilde{n}^2}\right),
	\end{equation}
	Using $\alpha\simeq7.5\times10^9(M/M_\odot)(\mu/1\text{eV})$, where $M_\odot$ is the solar mass, we have
	\begin{equation}
		f_{\tilde{n}\tilde{m}}\simeq 43\text{Hz}\left(\frac{M}{M_\odot}\right)^2\left(\frac{\mu}{10^{-11}\text{eV}}\right)^3\left(\frac{1}{\tilde{m}^2}-\frac{1}{\tilde{n}^2}\right),
	\end{equation}
	which falls in the sensitivity band of advanced ground based detectors for $\mu\sim10^{-11}\text{eV}$ and stellar-mass black holes \cite{ligocaltech}, and of space based detector for $\mu\sim10^{-15}\text{eV}$ and
	super massive black holes $\left(10^6-10^{8}\;M_{\odot}\right)$ \cite{lisa,Hu:2017mde,Hu:2018yqb}. The bound scalar particles will all be in ground state if there is no extra perturbations since they are bosons. With future
	developments, a careful analysis of the stochastic gravitational wave
	background \cite{Romano:2019yrj} might point to absorption lines present in it. This could
	be used to identify sources of the type described in this paper. In
	analogy with the quantization of electromagnetic waves from
	photoelectric experiments, one might get from this analysis a glimpse
	of the quantization of gravitational waves.
	
	\subsection{Radial distribution}
\begin{figure*}
		\centering	
		\includegraphics[width=0.32\textwidth,height=0.24\textwidth]{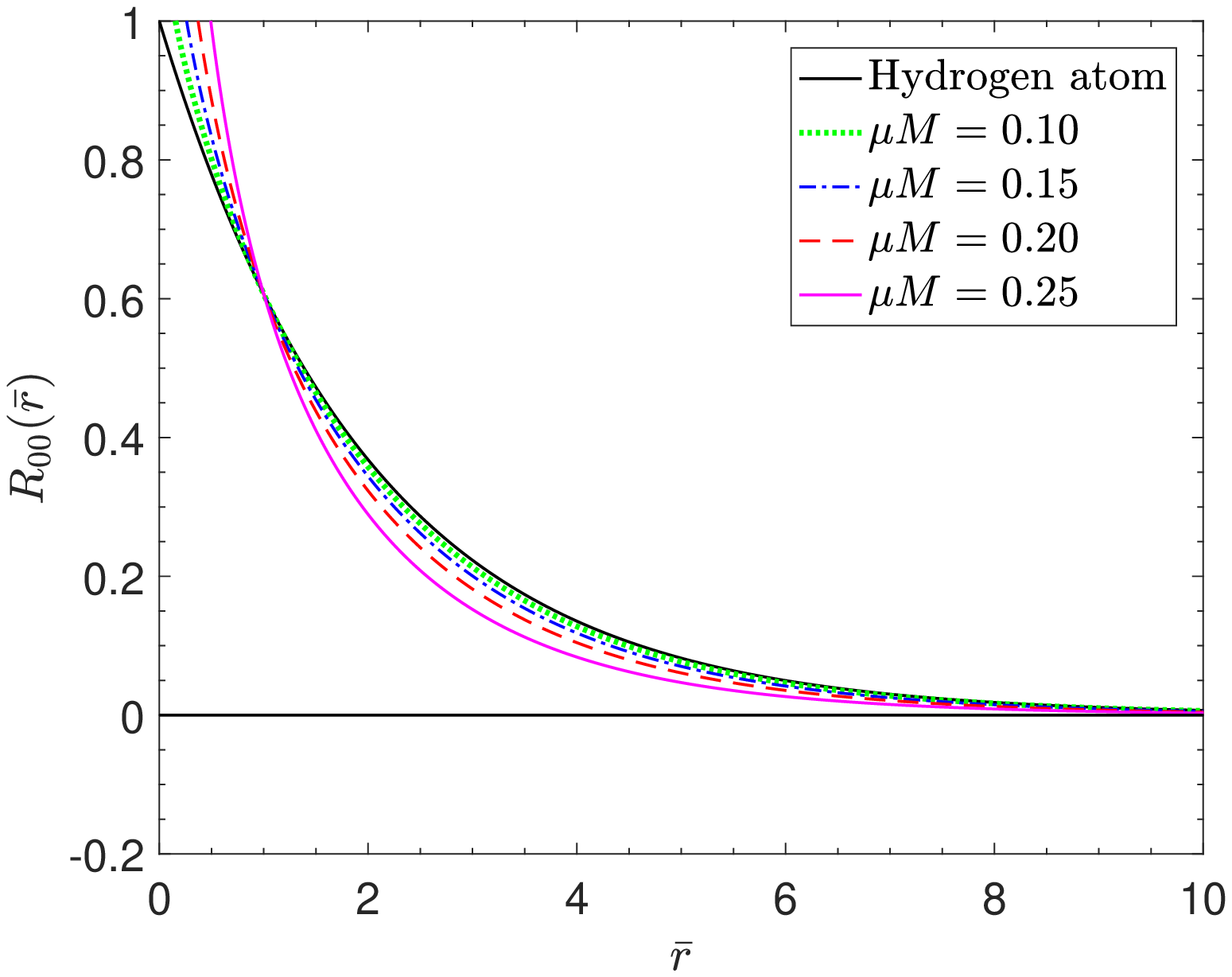}		
		\includegraphics[width=0.32\textwidth,height=0.24\textwidth]{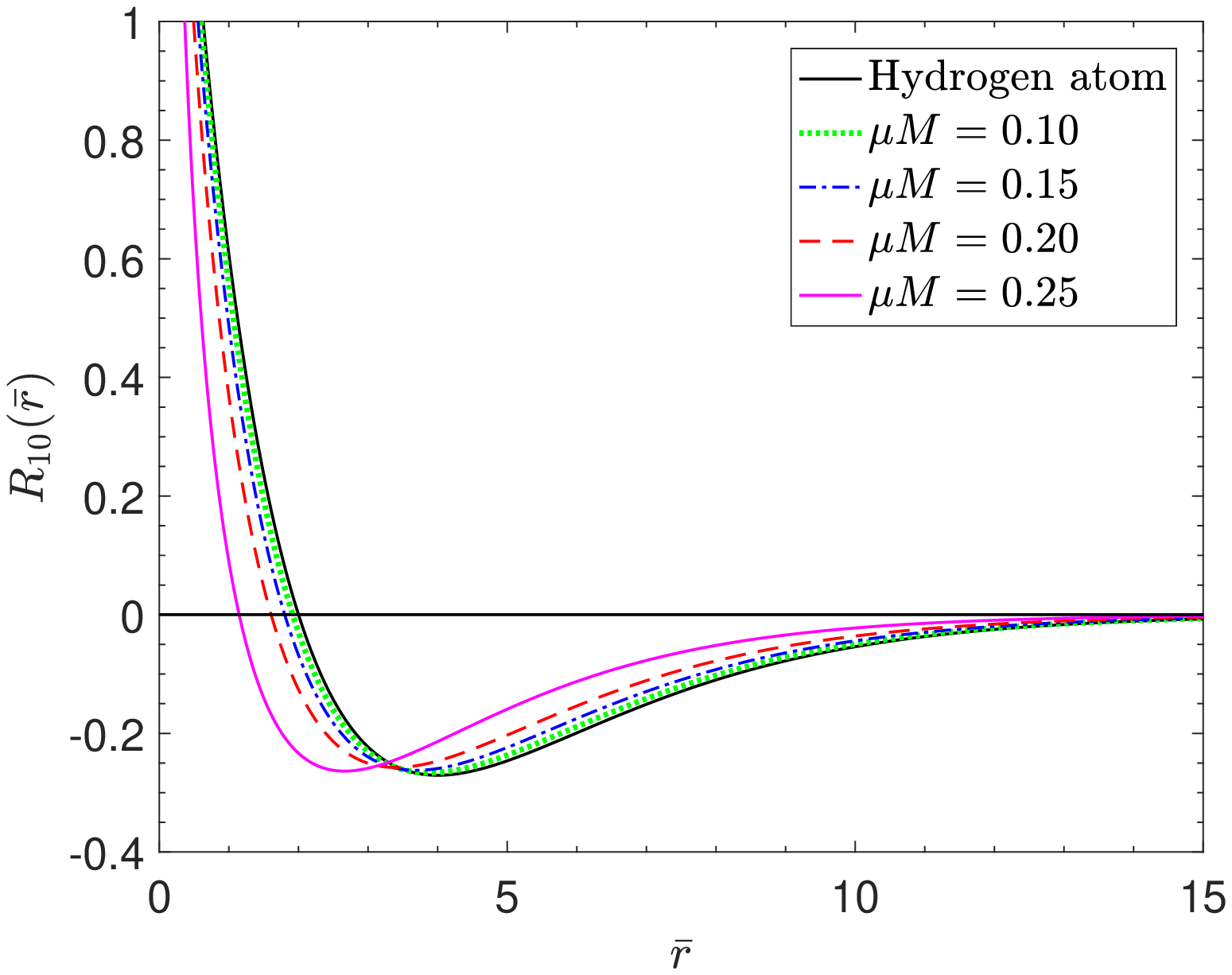}	
		\includegraphics[width=0.32\textwidth,height=0.24\textwidth]{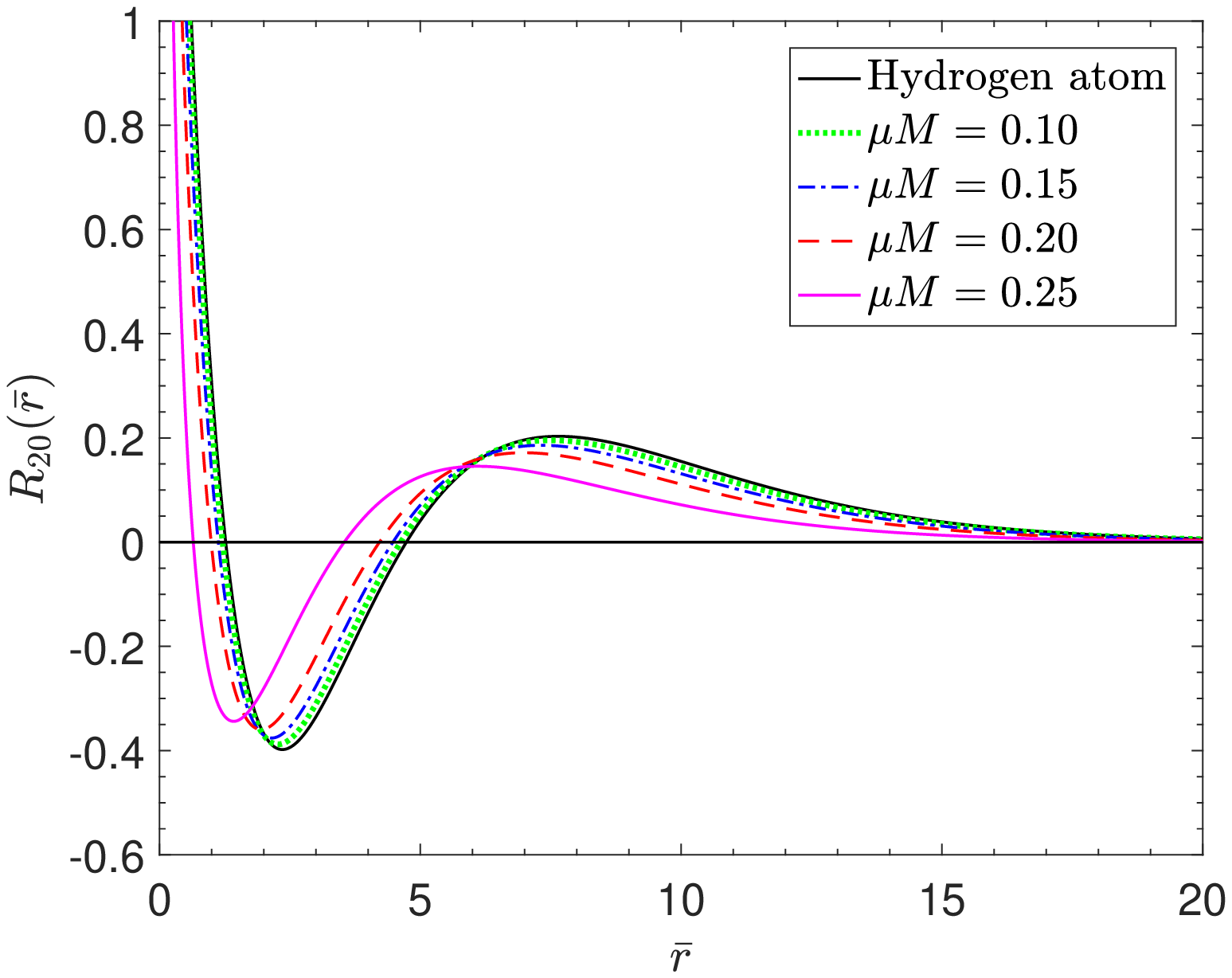}
		
		\includegraphics[width=0.32\textwidth,height=0.24\textwidth]{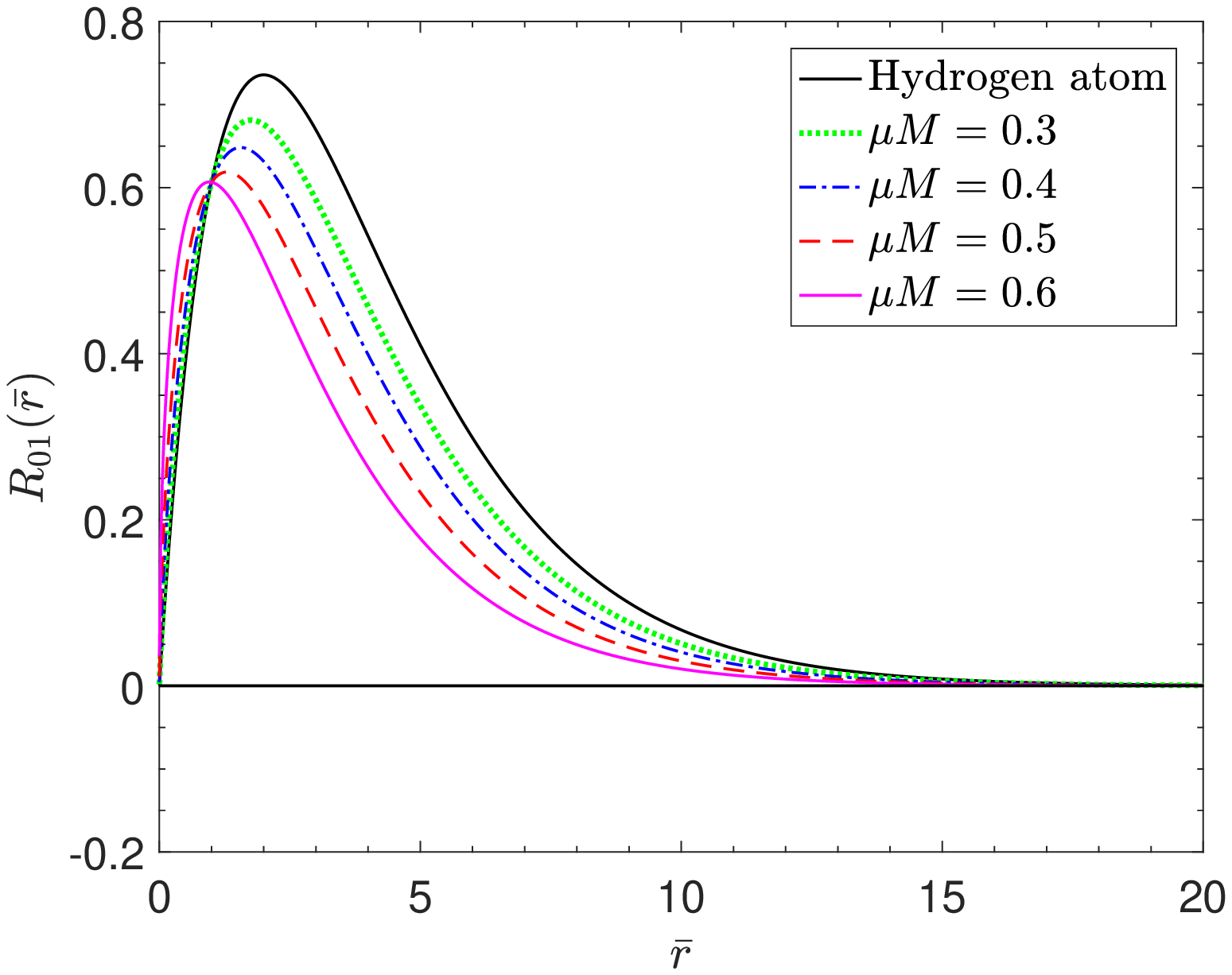}		
		\includegraphics[width=0.32\textwidth,height=0.24\textwidth]{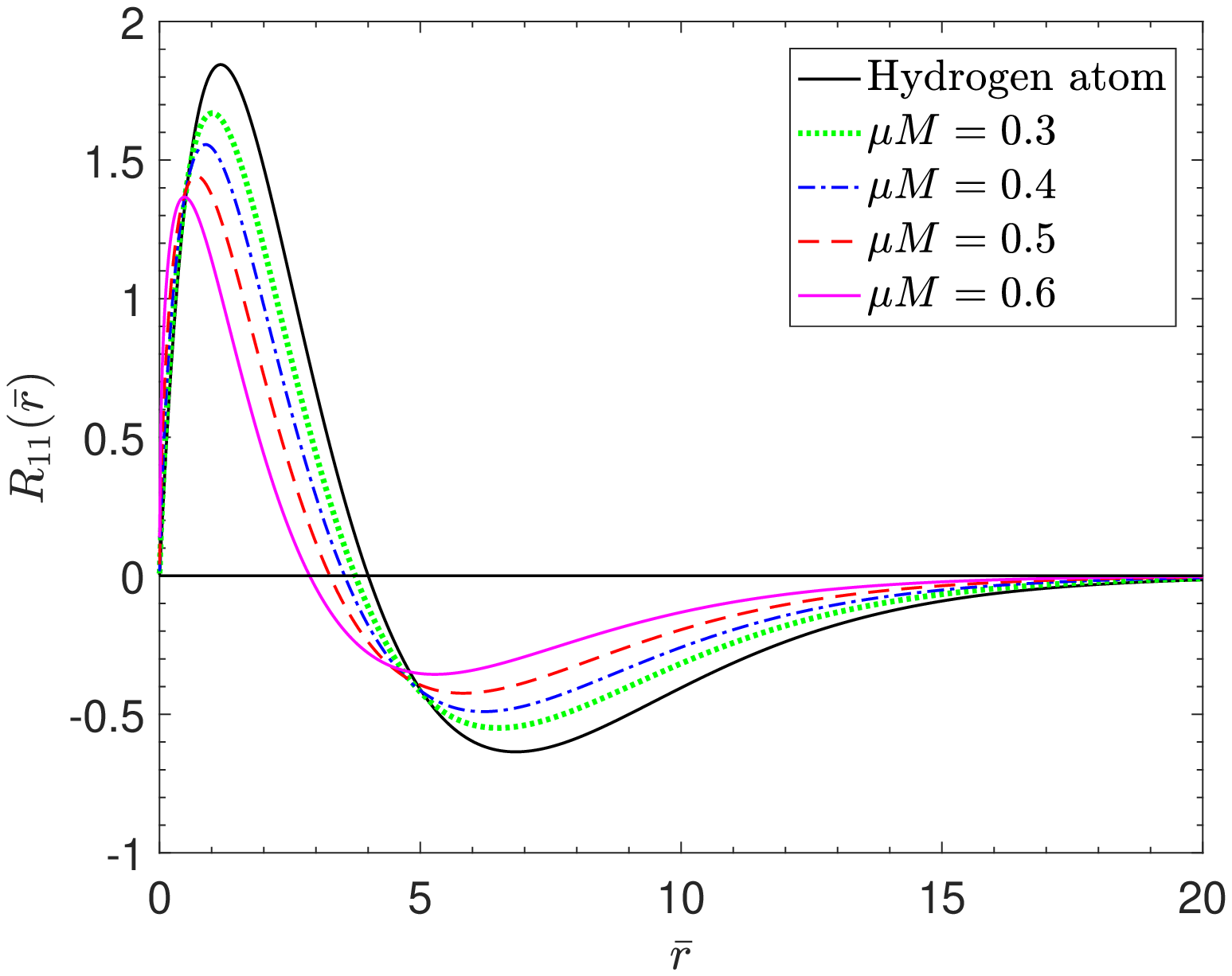}	
		\includegraphics[width=0.32\textwidth,height=0.24\textwidth]{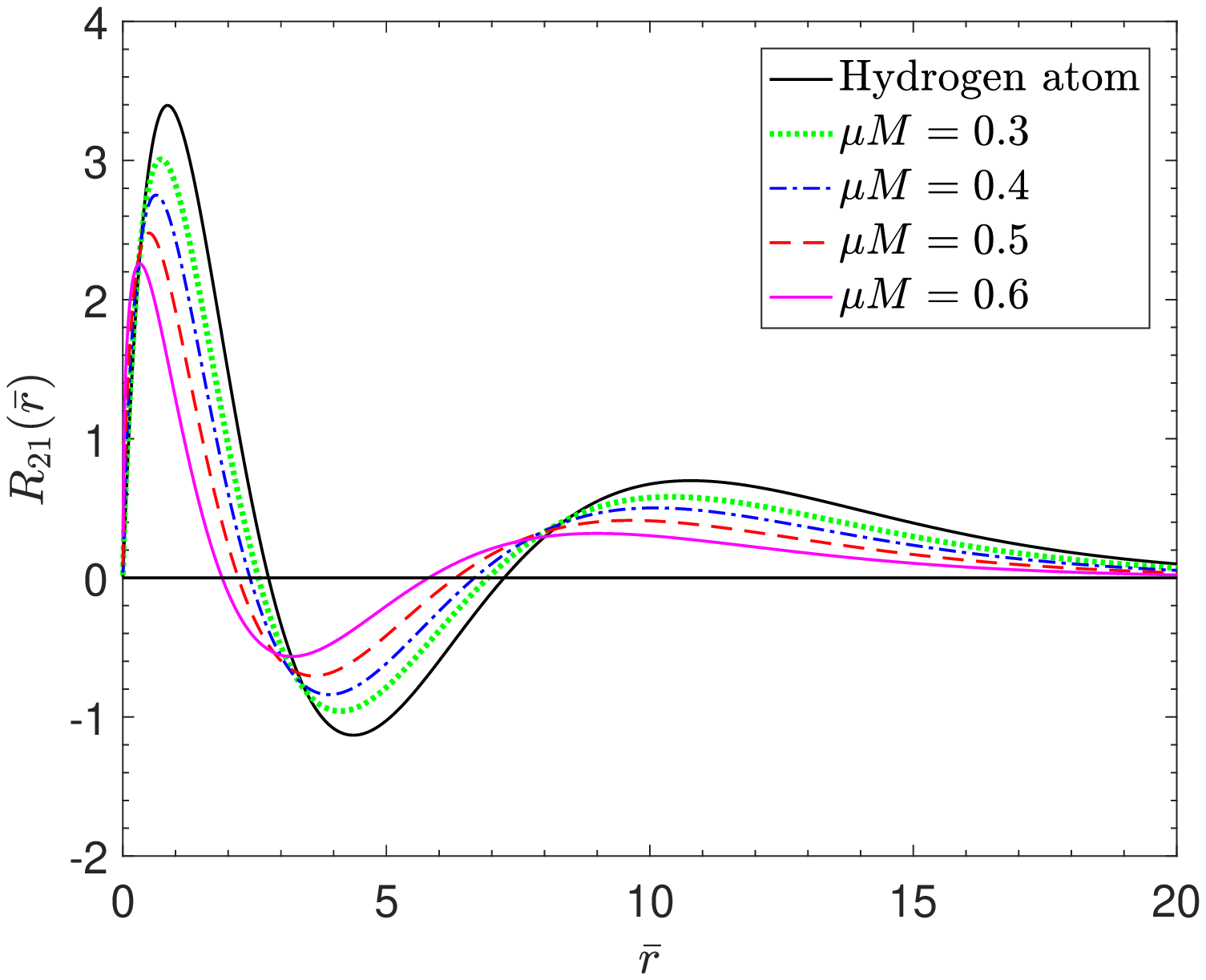}
		\caption{Comparison of radial functions given in (\ref{Eq: radial func n}), for values of $\mu M$.
			The upper panel shows the $l=0$ states, whereas the bottom panel shows the $l=1$ states. $n=0,1,2$ from left to right panels.}
		\label{Fig: radial func l01}
	\end{figure*}

	Let us now investigate the spatial distribution of the scalar clouds. From Eq.(\ref{Eq: separate variable}), the radial part of the scalar field is given by $R(r)=\psi(r)/r\sqrt{G(r)}$. To compare with the hydrogen atom, we introduce a new variable $\bar{r}=2kz$, and label the radial function by quantum numbers $(n,l)$. Then, using Eqs.(\ref{Eq: radial func}) and (\ref{Eq: matching condition}), we obtain
	\begin{equation}\label{Eq: radial func n}
		R_{nl}(\bar{r})\propto\bar{r}^{\beta-1/2}e^{-\bar{r}/2}L_{n}^{2\beta}(\bar{r}),
	\end{equation}
	where $\beta$ is given in Eq.(\ref{Eq: Def of beta}). In the limit $\alpha\ll L$,
	\begin{equation}\label{Eq: approx of beta}
		\beta=l+\frac{1}{2}-\frac{4\alpha^2}{1+2l}-\mathcal{O}\left(\alpha^4\right).
	\end{equation}
	If we neglect the term containing $\alpha^2$, the radial function is hydrogenic-like
	\begin{equation}
		R_{nl}(\bar{r})\propto\bar{r}^{l}e^{-\bar{r}/2}L_{n}^{2l+1}(\bar{r}).
	\end{equation}
	Typical radial functions of the scalar clouds are presented in Fig. \ref{Fig: radial func l01}. We see that the overtone $n$ is exactly the number of nodes of $R_{nl}$, and $R_{nl}$ decrease to zero at spatial infinity. We also see that as $\mu M$ is decreased, the maximum of $R_{nl}$ is pushed far away from the black hole. This is consistent with Fig. \ref{Fig: potential extreme massive} that the local minimum of the potential well moves far way from the black hole with the decrease of $\mu M$.
	
	It should be noted that for $l>1$, we have $\beta>1/2$ (see Eq.(\ref{Eq: Def of beta})), and the radial function $R_{nl}(\bar{r})$ is always finite for $\bar{r}\in[0,+\infty)$. However, for $l=0$ state, $\beta<1/2$, and the radial function is divergent at the event horizon $\bar{r}=0$. Fortunately, the integral over $\bar{r}$ of the energy density remains finite. This can be seen in the following.
	
	Substituting the decomposition in spherical harmonics (\ref{Eq: separate variable}) into Eq.(\ref{Eq: Def of E}) and integrating over the sphere, we obtain $E=\sum_{l}E_{l}$, with
	\begin{equation}
		E_l=\int_{r_+}^{\infty}\rho_l(r)r^2dr,
	\end{equation}
	where
	\begin{equation}\label{Eq: Def of rho}
		\begin{aligned}
			\rho_l(r)=&-G\int T^0_{\;\;0} \sin\vartheta d\vartheta d\varphi\\
			=&g^{11}G\bigg|\frac{dR_{nl}}{dr}\bigg|^2+\left[\frac{l(l+1)}{r^2}+G\left(\mu^2-g^{00}\omega_{\tilde{n}}^2\right)\right]|R_{nl}|^2.
		\end{aligned}
	\end{equation}	
	The integral can also be evaluated in the dimensionless coordinate $\bar{r}=2k(r-r_+)/r_+$
	\begin{equation}\label{Eq: energy l}
		E_{l}=\int_{0}^{\infty}\mathcal{E}_l(\bar{r})d\bar{r}.
	\end{equation}
	The radial energy density of the scalar cloud is defined as the integrand of $E_l$, i.e., $\mathcal{E}_l(\bar{r})\propto\rho_l(r)r^2$.
	For $l\geq1$, both the radial function and energy density are regular everywhere. However, for $l=0$, they may be divergent at the horizon $\bar{r}=0$. Using Eqs.(\ref{Eq: radial func n})-(\ref{Eq: energy l}) and taking $l=0$, we have
	\begin{equation}
		\mathcal{E}_{l=0}(\bar{r})\sim\bar{r}^{-8\alpha^2-\mathcal{O}\left(\alpha^4\right)},\;\;\;\text{for}\;\;\bar{r}\rightarrow0.
	\end{equation}
	The potential analysis in previous section shows that for large values of $\alpha$, the potential well may disappear, and no bound states exist. Conversely, for $\alpha<L/2$, the power of $\bar{r}$ in the energy density $\mathcal{E}_{l=0}(\bar{r})$ is greater than $-1$. Therefore, the energy given in Eq.(\ref{Eq: energy l}) is finite for $l=0$ in spite of the divergence of $\mathcal{E}_{l=0}(\bar{r})$ at $\bar{r}=0$.
	
\begin{figure}
		\centering		
		\includegraphics[width=0.45\textwidth,height=0.35\textwidth]{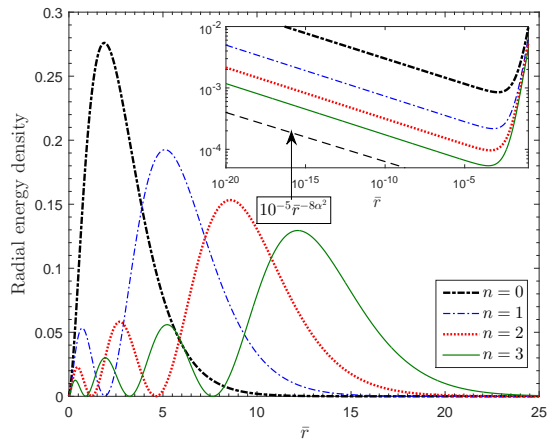}		
		\caption{Comparison of the radial energy densities, with $l=0$ and $\alpha=\mu M=0.1$, for $n=0,1,2,3$. All the energy densities have been normalized by $E_l=1$.}
		\label{Fig: radial energy density1}
	\end{figure}

	Figure \ref{Fig: radial energy density1} compares the radial energy densities of $l=0$ states for different values of $n$. We see that these energy densities are divergent as $\bar{r}\rightarrow0$. However, the integral given in Eq.(\ref{Eq: energy l}) remains finite, and the energy densities in Fig.\ref{Fig: radial energy density1} have been normalized by $E_{l=0}=1$. We also see that for a given $n$, the energy density is always positive, and the number of the maximum points is given by $n+1$. Like the radial function, the density decreases to zero exponentially at large distances.
	
	\section{Discussion and Conclusion}\label{Sec: conclusion}
	Black hole takes a pivotal status in any forms of quantization of gravity. An ordinary black hole may be in highly excited state in loop quantum gravity \cite{Perez:2017cmj}. String theory has a similar view about a black
	hole \cite{doi:10.1146/annurev.nucl.50.1.153}. Generally, ground state is much simpler than excited state to explore. Thus, one must be curious that is there a black hole happens to fall into ground state?  The property of
	zero-entropy of an extreme GMGHS black hole implies that it may be in ground state.  In this paper, we demonstrate that extreme GMGHS black hole really supports stable scalar clouds in its exterior spacetime, following a true atom.

	First, we explore the effective potential and find its width goes to infinity in tortoise coordinates when the GMGHS black hole becomes extremal. And then we find the exact solution of the radial function and present the energy spectrum of the bound states. The energy spectrum is pure real, since the particles around the hole stop penetrating into the hole, which are blocked by the infinite wide potential barrier. Further, we investigate the energy flux of the surrounding particles.  We show that the flux vanishes at the horizon, which confirms that the penetrating processes of particles from the exterior into the inner space of the hole stop.
	
	Through the above studies, we prove that a stable cloud can be formed surrounding an extreme GMGHS black hole. Some interesting properties of the scalar clouds are discussed. First, we show that these clouds are described by real bound states with $\omega_I=0$, and their frequencies are exactly given by the extreme limit of the quasibound state spectrum \cite{PhysRevD.103.044062}. The mechanism of the scalar cloud in present work does not depend on superradiance, and also applies to massive Dirac field in the GMGHS spacetime \cite{Huang2022}. Second, the radial function expressed in terms of the generalized Laguerre polynomials, and the quantum number $n$ indicates the nodes of the radial functions. We present a formula similar to  Ritz’s principle in Hydrogen atom theory.  More properties of extreme black holes in analogy to an atom need to explore further.

	\begin{acknowledgments}
		Our thanks go to the anonymous referees who present several valuable suggestions. This work is supported by  Shandong Province Natural Science Foundation under grant No. ZR201709220395.
	\end{acknowledgments}

	\bibliography{cloud}

\begin{thebibliography}{27}%
\makeatletter
\providecommand \@ifxundefined [1]{%
 \@ifx{#1\undefined}
}%
\providecommand \@ifnum [1]{%
 \ifnum #1\expandafter \@firstoftwo
 \else \expandafter \@secondoftwo
 \fi
}%
\providecommand \@ifx [1]{%
 \ifx #1\expandafter \@firstoftwo
 \else \expandafter \@secondoftwo
 \fi
}%
\providecommand \natexlab [1]{#1}%
\providecommand \enquote  [1]{``#1''}%
\providecommand \bibnamefont  [1]{#1}%
\providecommand \bibfnamefont [1]{#1}%
\providecommand \citenamefont [1]{#1}%
\providecommand \href@noop [0]{\@secondoftwo}%
\providecommand \href [0]{\begingroup \@sanitize@url \@href}%
\providecommand \@href[1]{\@@startlink{#1}\@@href}%
\providecommand \@@href[1]{\endgroup#1\@@endlink}%
\providecommand \@sanitize@url [0]{\catcode `\\12\catcode `\$12\catcode
  `\&12\catcode `\#12\catcode `\^12\catcode `\_12\catcode `\%12\relax}%
\providecommand \@@startlink[1]{}%
\providecommand \@@endlink[0]{}%
\providecommand \url  [0]{\begingroup\@sanitize@url \@url }%
\providecommand \@url [1]{\endgroup\@href {#1}{\urlprefix }}%
\providecommand \urlprefix  [0]{URL }%
\providecommand \Eprint [0]{\href }%
\providecommand \doibase [0]{http://dx.doi.org/}%
\providecommand \selectlanguage [0]{\@gobble}%
\providecommand \bibinfo  [0]{\@secondoftwo}%
\providecommand \bibfield  [0]{\@secondoftwo}%
\providecommand \translation [1]{[#1]}%
\providecommand \BibitemOpen [0]{}%
\providecommand \bibitemStop [0]{}%
\providecommand \bibitemNoStop [0]{.\EOS\space}%
\providecommand \EOS [0]{\spacefactor3000\relax}%
\providecommand \BibitemShut  [1]{\csname bibitem#1\endcsname}%
\let\auto@bib@innerbib\@empty
\bibitem [{\citenamefont {Detweiler}(1980)}]{Detweiler:1980uk}%
  \BibitemOpen
  \bibfield  {author} {\bibinfo {author} {\bibfnamefont {S.~L.}\ \bibnamefont
  {Detweiler}},\ }\href {\doibase 10.1103/PhysRevD.22.2323} {\bibfield
  {journal} {\bibinfo  {journal} {Phys. Rev. D}\ }\textbf {\bibinfo {volume}
  {22}},\ \bibinfo {pages} {2323} (\bibinfo {year} {1980})}\BibitemShut
  {NoStop}%
\bibitem [{\citenamefont {Nielsen}\ \emph {et~al.}(2019)\citenamefont
  {Nielsen}, \citenamefont {Palessandro},\ and\ \citenamefont
  {Sloth}}]{PhysRevD.99.123011}%
  \BibitemOpen
  \bibfield  {author} {\bibinfo {author} {\bibfnamefont {N.~G.}\ \bibnamefont
  {Nielsen}}, \bibinfo {author} {\bibfnamefont {A.}~\bibnamefont
  {Palessandro}}, \ and\ \bibinfo {author} {\bibfnamefont {M.~S.}\ \bibnamefont
  {Sloth}},\ }\href {\doibase 10.1103/PhysRevD.99.123011} {\bibfield  {journal}
  {\bibinfo  {journal} {Phys. Rev. D}\ }\textbf {\bibinfo {volume} {99}},\
  \bibinfo {pages} {123011} (\bibinfo {year} {2019})}\BibitemShut {NoStop}%
\bibitem [{\citenamefont {Starobinsky}(1973)}]{Starobinsky:1973aij}%
  \BibitemOpen
  \bibfield  {author} {\bibinfo {author} {\bibfnamefont {A.~A.}\ \bibnamefont
  {Starobinsky}},\ }\href@noop {} {\bibfield  {journal} {\bibinfo  {journal}
  {Sov. Phys. JETP}\ }\textbf {\bibinfo {volume} {37}},\ \bibinfo {pages} {28}
  (\bibinfo {year} {1973})}\BibitemShut {NoStop}%
\bibitem [{\citenamefont {Unruh}(1974)}]{PhysRevD.10.3194}%
  \BibitemOpen
  \bibfield  {author} {\bibinfo {author} {\bibfnamefont {W.~G.}\ \bibnamefont
  {Unruh}},\ }\href {\doibase 10.1103/PhysRevD.10.3194} {\bibfield  {journal}
  {\bibinfo  {journal} {Phys. Rev. D}\ }\textbf {\bibinfo {volume} {10}},\
  \bibinfo {pages} {3194} (\bibinfo {year} {1974})}\BibitemShut {NoStop}%
\bibitem [{\citenamefont {Brito}\ \emph {et~al.}(2015)\citenamefont {Brito},
  \citenamefont {Cardoso},\ and\ \citenamefont {Pani}}]{Brito:2015oca}%
  \BibitemOpen
  \bibfield  {author} {\bibinfo {author} {\bibfnamefont {R.}~\bibnamefont
  {Brito}}, \bibinfo {author} {\bibfnamefont {V.}~\bibnamefont {Cardoso}}, \
  and\ \bibinfo {author} {\bibfnamefont {P.}~\bibnamefont {Pani}},\ }\href
  {\doibase 10.1007/978-3-319-19000-6} {\bibfield  {journal} {\bibinfo
  {journal} {Lect. Notes Phys.}\ }\textbf {\bibinfo {volume} {906}},\ \bibinfo
  {pages} {pp.1} (\bibinfo {year} {2015})},\ \Eprint
  {http://arxiv.org/abs/1501.06570} {arXiv:1501.06570 [gr-qc]} \BibitemShut
  {NoStop}%
\bibitem [{\citenamefont {Hod}(2012)}]{PhysRevD.86.104026}%
  \BibitemOpen
  \bibfield  {author} {\bibinfo {author} {\bibfnamefont {S.}~\bibnamefont
  {Hod}},\ }\href {\doibase 10.1103/PhysRevD.86.104026} {\bibfield  {journal}
  {\bibinfo  {journal} {Phys. Rev. D}\ }\textbf {\bibinfo {volume} {86}},\
  \bibinfo {pages} {104026} (\bibinfo {year} {2012})}\BibitemShut {NoStop}%
\bibitem [{\citenamefont {Huang}\ and\ \citenamefont
  {Liu}(2016)}]{Huang:2016qnk}%
  \BibitemOpen
  \bibfield  {author} {\bibinfo {author} {\bibfnamefont {Y.}~\bibnamefont
  {Huang}}\ and\ \bibinfo {author} {\bibfnamefont {D.-J.}\ \bibnamefont
  {Liu}},\ }\href {\doibase 10.1103/PhysRevD.94.064030} {\bibfield  {journal}
  {\bibinfo  {journal} {Phys. Rev. D}\ }\textbf {\bibinfo {volume} {94}},\
  \bibinfo {pages} {064030} (\bibinfo {year} {2016})},\ \Eprint
  {http://arxiv.org/abs/1606.08913} {arXiv:1606.08913 [gr-qc]} \BibitemShut
  {NoStop}%
\bibitem [{\citenamefont {Huang}\ \emph {et~al.}(2017)\citenamefont {Huang},
  \citenamefont {Liu}, \citenamefont {Zhai},\ and\ \citenamefont
  {Li}}]{Huang:2017whw}%
  \BibitemOpen
  \bibfield  {author} {\bibinfo {author} {\bibfnamefont {Y.}~\bibnamefont
  {Huang}}, \bibinfo {author} {\bibfnamefont {D.-J.}\ \bibnamefont {Liu}},
  \bibinfo {author} {\bibfnamefont {X.-H.}\ \bibnamefont {Zhai}}, \ and\
  \bibinfo {author} {\bibfnamefont {X.-Z.}\ \bibnamefont {Li}},\ }\href
  {\doibase 10.1088/1361-6382/aa7964} {\bibfield  {journal} {\bibinfo
  {journal} {Class. Quant. Grav.}\ }\textbf {\bibinfo {volume} {34}},\ \bibinfo
  {pages} {155002} (\bibinfo {year} {2017})},\ \Eprint
  {http://arxiv.org/abs/1706.04441} {arXiv:1706.04441 [gr-qc]} \BibitemShut
  {NoStop}%
\bibitem [{\citenamefont {Herdeiro}\ and\ \citenamefont
  {Radu}(2014)}]{Herdeiro:2014goa}%
  \BibitemOpen
  \bibfield  {author} {\bibinfo {author} {\bibfnamefont {C.~A.~R.}\
  \bibnamefont {Herdeiro}}\ and\ \bibinfo {author} {\bibfnamefont
  {E.}~\bibnamefont {Radu}},\ }\href {\doibase 10.1103/PhysRevLett.112.221101}
  {\bibfield  {journal} {\bibinfo  {journal} {Phys. Rev. Lett.}\ }\textbf
  {\bibinfo {volume} {112}},\ \bibinfo {pages} {221101} (\bibinfo {year}
  {2014})},\ \Eprint {http://arxiv.org/abs/1403.2757} {arXiv:1403.2757 [gr-qc]}
  \BibitemShut {NoStop}%
\bibitem [{\citenamefont {Ganchev}\ and\ \citenamefont
  {Santos}(2018)}]{PhysRevLett.120.171101}%
  \BibitemOpen
  \bibfield  {author} {\bibinfo {author} {\bibfnamefont {B.}~\bibnamefont
  {Ganchev}}\ and\ \bibinfo {author} {\bibfnamefont {J.~E.}\ \bibnamefont
  {Santos}},\ }\href {\doibase 10.1103/PhysRevLett.120.171101} {\bibfield
  {journal} {\bibinfo  {journal} {Phys. Rev. Lett.}\ }\textbf {\bibinfo
  {volume} {120}},\ \bibinfo {pages} {171101} (\bibinfo {year}
  {2018})}\BibitemShut {NoStop}%
\bibitem [{\citenamefont {Holzhey}\ and\ \citenamefont
  {Wilczek}(1992)}]{Holzhey:1991bx}%
  \BibitemOpen
  \bibfield  {author} {\bibinfo {author} {\bibfnamefont {C.~F.~E.}\
  \bibnamefont {Holzhey}}\ and\ \bibinfo {author} {\bibfnamefont
  {F.}~\bibnamefont {Wilczek}},\ }\href {\doibase 10.1016/0550-3213(92)90254-9}
  {\bibfield  {journal} {\bibinfo  {journal} {Nucl. Phys. B}\ }\textbf
  {\bibinfo {volume} {380}},\ \bibinfo {pages} {447} (\bibinfo {year}
  {1992})},\ \Eprint {http://arxiv.org/abs/hep-th/9202014}
  {arXiv:hep-th/9202014} \BibitemShut {NoStop}%
\bibitem [{\citenamefont {Preskill}\ \emph {et~al.}(1991)\citenamefont
  {Preskill}, \citenamefont {Schwarz}, \citenamefont {Shapere}, \citenamefont
  {Trivedi},\ and\ \citenamefont {Wilczek}}]{doi:10.1142/S0217732391002773}%
  \BibitemOpen
  \bibfield  {author} {\bibinfo {author} {\bibfnamefont {J.}~\bibnamefont
  {Preskill}}, \bibinfo {author} {\bibfnamefont {P.}~\bibnamefont {Schwarz}},
  \bibinfo {author} {\bibfnamefont {A.}~\bibnamefont {Shapere}}, \bibinfo
  {author} {\bibfnamefont {S.}~\bibnamefont {Trivedi}}, \ and\ \bibinfo
  {author} {\bibfnamefont {F.}~\bibnamefont {Wilczek}},\ }\href {\doibase
  10.1142/S0217732391002773} {\bibfield  {journal} {\bibinfo  {journal} {Modern
  Physics Letters A}\ }\textbf {\bibinfo {volume} {06}},\ \bibinfo {pages}
  {2353} (\bibinfo {year} {1991})}\BibitemShut {NoStop}%
\bibitem [{\citenamefont {Appelquist}\ \emph {et~al.}(1987)\citenamefont
  {Appelquist}, \citenamefont {Chodos},\ and\ \citenamefont
  {Freund}}]{Appelquist:1987nr}%
  \BibitemOpen
  \bibinfo {editor} {\bibfnamefont {T.}~\bibnamefont {Appelquist}}, \bibinfo
  {editor} {\bibfnamefont {A.}~\bibnamefont {Chodos}}, \ and\ \bibinfo {editor}
  {\bibfnamefont {P.~G.~O.}\ \bibnamefont {Freund}},\ eds.,\ \href@noop {}
  {\emph {\bibinfo {title} {{MODERN KALUZA-KLEIN THEORIES}}}}\ (\bibinfo {year}
  {1987})\BibitemShut {NoStop}%
\bibitem [{\citenamefont {Dirac}(1937)}]{Dirac:1937ti}%
  \BibitemOpen
  \bibfield  {author} {\bibinfo {author} {\bibfnamefont {P.~A.~M.}\
  \bibnamefont {Dirac}},\ }\href {\doibase 10.1038/139323a0} {\bibfield
  {journal} {\bibinfo  {journal} {Nature}\ }\textbf {\bibinfo {volume} {139}},\
  \bibinfo {pages} {323} (\bibinfo {year} {1937})}\BibitemShut {NoStop}%
\bibitem [{\citenamefont {Garfinkle}\ \emph {et~al.}(1991)\citenamefont
  {Garfinkle}, \citenamefont {Horowitz},\ and\ \citenamefont
  {Strominger}}]{Garfinkle:1990qj}%
  \BibitemOpen
  \bibfield  {author} {\bibinfo {author} {\bibfnamefont {D.}~\bibnamefont
  {Garfinkle}}, \bibinfo {author} {\bibfnamefont {G.~T.}\ \bibnamefont
  {Horowitz}}, \ and\ \bibinfo {author} {\bibfnamefont {A.}~\bibnamefont
  {Strominger}},\ }\href {\doibase 10.1103/PhysRevD.43.3140,
  10.1103/PhysRevD.45.3888} {\bibfield  {journal} {\bibinfo  {journal} {Phys.
  Rev.}\ }\textbf {\bibinfo {volume} {D43}},\ \bibinfo {pages} {3140} (\bibinfo
  {year} {1991})},\ \bibinfo {note} {[Erratum: Phys.
  Rev.D45,3888(1992)]}\BibitemShut {NoStop}%
\bibitem [{\citenamefont {Gibbons}\ and\ \citenamefont
  {Maeda}(1988)}]{Gibbons:1987ps}%
  \BibitemOpen
  \bibfield  {author} {\bibinfo {author} {\bibfnamefont {G.~W.}\ \bibnamefont
  {Gibbons}}\ and\ \bibinfo {author} {\bibfnamefont {K.-i.}\ \bibnamefont
  {Maeda}},\ }\href {\doibase 10.1016/0550-3213(88)90006-5} {\bibfield
  {journal} {\bibinfo  {journal} {Nucl. Phys.}\ }\textbf {\bibinfo {volume}
  {B298}},\ \bibinfo {pages} {741} (\bibinfo {year} {1988})}\BibitemShut
  {NoStop}%
\bibitem [{\citenamefont {Huang}\ and\ \citenamefont
  {Zhang}(2020)}]{Huang:2020bdf}%
  \BibitemOpen
  \bibfield  {author} {\bibinfo {author} {\bibfnamefont {Y.}~\bibnamefont
  {Huang}}\ and\ \bibinfo {author} {\bibfnamefont {H.}~\bibnamefont {Zhang}},\
  }\href {\doibase 10.1140/epjc/s10052-020-8228-8} {\bibfield  {journal}
  {\bibinfo  {journal} {Eur. Phys. J. C}\ }\textbf {\bibinfo {volume} {80}},\
  \bibinfo {pages} {654} (\bibinfo {year} {2020})},\ \Eprint
  {http://arxiv.org/abs/2006.01388} {arXiv:2006.01388 [gr-qc]} \BibitemShut
  {NoStop}%
\bibitem [{\citenamefont {Huang}\ and\ \citenamefont
  {Zhang}(2021)}]{PhysRevD.103.044062}%
  \BibitemOpen
  \bibfield  {author} {\bibinfo {author} {\bibfnamefont {Y.}~\bibnamefont
  {Huang}}\ and\ \bibinfo {author} {\bibfnamefont {H.}~\bibnamefont {Zhang}},\
  }\href {\doibase 10.1103/PhysRevD.103.044062} {\bibfield  {journal} {\bibinfo
   {journal} {Phys. Rev. D}\ }\textbf {\bibinfo {volume} {103}},\ \bibinfo
  {pages} {044062} (\bibinfo {year} {2021})}\BibitemShut {NoStop}%
\bibitem [{\citenamefont {Degollado}\ and\ \citenamefont
  {Herdeiro}(2014)}]{Degollado:2013bha}%
  \BibitemOpen
  \bibfield  {author} {\bibinfo {author} {\bibfnamefont {J.~C.}\ \bibnamefont
  {Degollado}}\ and\ \bibinfo {author} {\bibfnamefont {C.~A.~R.}\ \bibnamefont
  {Herdeiro}},\ }\href {\doibase 10.1103/PhysRevD.89.063005} {\bibfield
  {journal} {\bibinfo  {journal} {Phys. Rev. D}\ }\textbf {\bibinfo {volume}
  {89}},\ \bibinfo {pages} {063005} (\bibinfo {year} {2014})},\ \Eprint
  {http://arxiv.org/abs/1312.4579} {arXiv:1312.4579 [gr-qc]} \BibitemShut
  {NoStop}%
\bibitem [{lig()}]{ligocaltech}%
  \BibitemOpen
  \href@noop {} {}\bibinfo {howpublished}
  {https://www.ligo.caltech.edu/}\BibitemShut {NoStop}%
\bibitem [{lis()}]{lisa}%
  \BibitemOpen
  \href@noop {} {}\bibinfo {howpublished}
  {https://lisa.nasa.gov/index.html}\BibitemShut {NoStop}%
\bibitem [{\citenamefont {Hu}\ and\ \citenamefont {Wu}(2017)}]{Hu:2017mde}%
  \BibitemOpen
  \bibfield  {author} {\bibinfo {author} {\bibfnamefont {W.-R.}\ \bibnamefont
  {Hu}}\ and\ \bibinfo {author} {\bibfnamefont {Y.-L.}\ \bibnamefont {Wu}},\
  }\href {\doibase 10.1093/nsr/nwx116} {\bibfield  {journal} {\bibinfo
  {journal} {Natl. Sci. Rev.}\ }\textbf {\bibinfo {volume} {4}},\ \bibinfo
  {pages} {685} (\bibinfo {year} {2017})}\BibitemShut {NoStop}%
\bibitem [{\citenamefont {Hu}\ \emph {et~al.}(2018)\citenamefont {Hu},
  \citenamefont {Li}, \citenamefont {Wang}, \citenamefont {Feng}, \citenamefont
  {Zhou}, \citenamefont {Hu}, \citenamefont {Hu}, \citenamefont {Mei},\ and\
  \citenamefont {Shao}}]{Hu:2018yqb}%
  \BibitemOpen
  \bibfield  {author} {\bibinfo {author} {\bibfnamefont {X.-C.}\ \bibnamefont
  {Hu}}, \bibinfo {author} {\bibfnamefont {X.-H.}\ \bibnamefont {Li}}, \bibinfo
  {author} {\bibfnamefont {Y.}~\bibnamefont {Wang}}, \bibinfo {author}
  {\bibfnamefont {W.-F.}\ \bibnamefont {Feng}}, \bibinfo {author}
  {\bibfnamefont {M.-Y.}\ \bibnamefont {Zhou}}, \bibinfo {author}
  {\bibfnamefont {Y.-M.}\ \bibnamefont {Hu}}, \bibinfo {author} {\bibfnamefont
  {S.-C.}\ \bibnamefont {Hu}}, \bibinfo {author} {\bibfnamefont {J.-W.}\
  \bibnamefont {Mei}}, \ and\ \bibinfo {author} {\bibfnamefont {C.-G.}\
  \bibnamefont {Shao}},\ }\href {\doibase 10.1088/1361-6382/aab52f} {\bibfield
  {journal} {\bibinfo  {journal} {Class. Quant. Grav.}\ }\textbf {\bibinfo
  {volume} {35}},\ \bibinfo {pages} {095008} (\bibinfo {year} {2018})},\
  \Eprint {http://arxiv.org/abs/1803.03368} {arXiv:1803.03368 [gr-qc]}
  \BibitemShut {NoStop}%
\bibitem [{\citenamefont {Romano}(2019)}]{Romano:2019yrj}%
  \BibitemOpen
  \bibfield  {author} {\bibinfo {author} {\bibfnamefont {J.~D.}\ \bibnamefont
  {Romano}}\ }(\bibinfo {year} {2019})\ \Eprint
  {http://arxiv.org/abs/1909.00269} {arXiv:1909.00269 [gr-qc]} \BibitemShut
  {NoStop}%
\bibitem [{\citenamefont {Perez}(2017)}]{Perez:2017cmj}%
  \BibitemOpen
  \bibfield  {author} {\bibinfo {author} {\bibfnamefont {A.}~\bibnamefont
  {Perez}},\ }\href {\doibase 10.1088/1361-6633/aa7e14} {\bibfield  {journal}
  {\bibinfo  {journal} {Rept. Prog. Phys.}\ }\textbf {\bibinfo {volume} {80}},\
  \bibinfo {pages} {126901} (\bibinfo {year} {2017})},\ \Eprint
  {http://arxiv.org/abs/1703.09149} {arXiv:1703.09149 [gr-qc]} \BibitemShut
  {NoStop}%
\bibitem [{\citenamefont {Das}\ and\ \citenamefont
  {Mathur}(2000)}]{doi:10.1146/annurev.nucl.50.1.153}%
  \BibitemOpen
  \bibfield  {author} {\bibinfo {author} {\bibfnamefont {S.~R.}\ \bibnamefont
  {Das}}\ and\ \bibinfo {author} {\bibfnamefont {S.~D.}\ \bibnamefont
  {Mathur}},\ }\href {\doibase 10.1146/annurev.nucl.50.1.153} {\bibfield
  {journal} {\bibinfo  {journal} {Annual Review of Nuclear and Particle
  Science}\ }\textbf {\bibinfo {volume} {50}},\ \bibinfo {pages} {153}
  (\bibinfo {year} {2000})}\BibitemShut {NoStop}%
\bibitem [{\citenamefont {Huang}\ and\ \citenamefont
  {Zhang}(2022)}]{Huang2022}%
  \BibitemOpen
  \bibfield  {author} {\bibinfo {author} {\bibfnamefont {Y.}~\bibnamefont
  {Huang}}\ and\ \bibinfo {author} {\bibfnamefont {H.}~\bibnamefont {Zhang}},\
  }\href {\doibase 10.34133/2022/9823274} {\bibfield  {journal} {\bibinfo
  {journal} {Research}\ }\textbf {\bibinfo {volume} {2022}},\ \bibinfo {pages}
  {Article ID 9823274} (\bibinfo {year} {2022})}\BibitemShut {NoStop}%
\end{thebibliography}%

\end{document}